\documentclass[aps,twocolumn,prb,nobalancelastpage,nofootinbib]{revtex4-2}
\usepackage[utf8]{inputenc}
\usepackage{graphics}     
\usepackage{graphicx}
\usepackage{caption}
\usepackage{url}          
\usepackage{bm}          
\usepackage{mathtools,amsmath,amssymb,lipsum,xcolor}
\usepackage{enumitem}

\newcommand{\roundbracket}[1]{\left(#1\right)}
\newcommand{\squarebracket}[1]{\left[#1\right]}
\newcommand{\otherbracket}[1]{\left \{#1\right\}}

\newcommand{\wideeq}[1]{\begin{widetext} \begin{equation} #1 \end{equation}\end{widetext}}

\begin{document}
\title{Dissipative systems fractionally coupled to a bath}
\author         {A. Vertessen$^{1}$, R.~C. Verstraten$^{1}$, C. Morais Smith$^{1}$}
\affiliation    {$^{1}$\mbox{Institute for Theoretical Physics, Utrecht University, Princetonplein 5, 3584CC Utrecht, The Netherlands}}
\date{July 20, 2023}

\begin{abstract}
Quantum diffusion is a major topic in condensed-matter physics, and the Caldeira-Leggett model has been one of the most successful approaches to study this phenomenon. Here, we generalize this model by coupling the bath to the system through a Weyl fractional derivative. The Weyl fractional Langevin equation is then derived without imposing a non-Ohmic macroscopic spectral function for the bath. By investigating the short- and long-time behavior of the mean squared displacement (MSD), we show that this model is able to describe a large variety of anomalous diffusion. Indeed, we find ballistic, sub-ballistic, and super-ballistic behavior for short times, whereas for long times we find saturation, and sub- and super-diffusion.
\end{abstract}



\maketitle
\section{Introduction}
Quantum diffusion~\cite{Bao2021, Kagan1992, KLEINERT20101, sq, gs, blackholes} has been at the attention of researchers for many years, posing the fundamental questions on the description of dissipation in open quantum systems. A more complete understanding of these mechanisms could lead to important breakthroughs in engineering devices that have a better energy efficiency. Despite the existence of several models, numerous unanswered questions remain. For instance, experimental evidence~\cite{anomalous, regner2013anomalous, banks2005anomalous} indicates that diffusion does not always conform to the linear behavior described by Einstein~\cite{einstein}, but instead exhibits a power-law scaling in the mean squared displacement (MSD), commonly known as \textit{anomalous diffusion}~\cite{METZLER20001,chen2006time}. One such example is quantum diffusion on a fractal~\cite{exp1}, where the exponent scales with the fractal dimension. Typically, models that describe anomalous diffusion assume a different environment compared to regular diffusion. Our aim here is to achieve a description of anomalous diffusion within the usual environment using instead a fractional interaction.

In order to build a dissipative model, the environment of a system should be included in the Hamiltonian, since energy is conserved in quantum mechanics. In the \textit{Caldeira-Leggett} model~\cite{caldeira_2014,weiss2012quantum, CLmodel,leggett1987dynamics,caldeira1981influence,ferialdi2017dissipation}, the system of interest, which we will take to be a single quantum particle, is coupled linearly with a thermal bath that consists of harmonic oscillators. Although this might seem arbitrary at first, it is a very good approximation for coupling a system to an in-equilibrium environment, since each degree of freedom of such an environment must oscillate around a local minimum. The benefit of this approximation is that the harmonic oscillators can be integrated out exactly, thus leading to an effective description of the particle. In doing so, one needs to know the spectral function of the bath, which is given by the imaginary part of the Fourier transform of the retarded dynamical susceptibility. This spectral function is, however, often difficult to measure experimentally, and is therefore often assumed to be linear (\textit{Ohmic}) for simplicity. This assumption on the reservoir leads to a Langevin equation, which describes linear diffusion. However, more recent works~\cite{e_lutz,timeglass,burov2008critical,robin} have shown that a power-law (\textit{non-Ohmic}) bath leads to a \textit{fractional Langevin equation}, which will typically have colored noise due to the fluctuation-dissipation theorem, but can also have white noise~\cite{timeglass}.

In the fractional Langevin equation, the first derivative friction term is replaced by a fractional derivative~\cite{oldham1974fractional,mainardi2018fractional}. These are operators which generalize the order of derivatives to be not just an integer number, like a first or second derivative, but any real (or even complex) number. There are, however, many different fractional derivatives, which are not all equivalent~\cite{robin, fractionaldiffrev, fracriesz}, and this can often lead to different results. The first main question then is: which fractional derivative should one use? For this, we have to realize that fractional derivatives are non-local operators, which means that it is important to specify the time domain. Refs.~\cite{e_lutz,timeglass} both use a Caputo derivative, which is only possible if we have a clear initialization time.  Since we are interested in dynamics, the boundaries of the action are taken from $-\infty$ to $+\infty$, which is an issue if we want to use Caputo derivatives inside the action. Therefore, we must take the domain of the fractional derivative to be the same as of the action. Furthermore, one would like to preserve the Fourier transform, since this has proven to be a powerful mathematical tool in physics. Hence, we would like a fractional derivative that is compatible with it. Because of these two reasons, we thus select the \textit{Weyl fractional derivative}~\cite{ferrari2018weyl,samko1993fractional,miller2006weyl}.

Here, we generalize the Caldeira-Leggett model by introducing a Weyl fractional derivative coupling term and show that it can describe a whole family of anomalous diffusion. Unlike previous works~\cite{robin, timeglass, e_lutz}, we derive a fractional Langevin equation \textit{without imposing a non-Ohmic macroscopic spectral function}. Motivated by the general velocity depended coupling in Ref.~\cite{ruward}, we consider a fractional Weyl derivative to couple the system to the bath. In order to study the Lagrangian, 
we first derive the \textit{Weyl fractional Euler-Lagrange equation} in analogy with Ref.~\cite{AGRAWAL2002368}. 
Then, we are able to derive the \textit{Weyl fractional Langevin equation} for an Ohmic bath, which we solve analytically using Fox H-functions. We calculate the asymptotic behavior of the MSD and show that it is comparable to previous results of a non-Ohmic Caldeira-Leggett model~\cite{e_lutz,barkai2001fractional,magdziarz2009fractional,razminia2014analysis,chen2010anomalous,tarasov2013review}.


The outline of this paper is the following: In Sec.~\ref{sec2} we introduce the Weyl fractional derivative because it is not well known. Then, in Sec.~\ref{sec3} we derive the Weyl fractional Langevin equation. Its solution for the free case is presented in Sec.~\ref{sec4}. In Sec.~\ref{sec5}, we investigate the asymptotic behavior of the MSD, and in Sec.~\ref{sec6} we present our conclusions and outlook.


\section{The Weyl fractional derivative}
\label{sec2}

Although the Weyl fractional derivative has been mentioned briefly in the literature (see Refs.~\cite{pseudodiff, METZLER20001, timeglass, robin, fracltwo, qm0982}), its mathematical exploration has been very limited, to the best of our knowledge. Therefore, we provide a summary of its main properties. Let $\alpha \in \mathbb{R}$, then
\begin{align}
&\prescript{W}{+}{\bm{D}_t^\alpha}f(t) = \mathcal{F}^{-1}\otherbracket{(i\omega)^\alpha\mathcal{F}\squarebracket{f(t);\omega}; t};\nonumber \\
&\prescript{W}{-}{\bm{D}_t^\alpha}f(t) = \mathcal{F}^{-1}\otherbracket{(-i\omega)^\alpha\mathcal{F}\squarebracket{f(t);\omega}; t}
\label{eq:def_weyl}
\end{align}
are called the \textit{Weyl and pseudo Weyl fractional derivative} of order $\alpha$, respectively\footnote{The used conventions for the Fourier transform and its inverse are \begin{align}
&\mathcal{F}\squarebracket{f(t); \omega} = \frac{1}{2\pi}\int_{-\infty}^\infty dt f(t) e^{-i\omega t}, \notag \\
&\mathcal{F}^{-1}\squarebracket{f(\omega); t} = \int_{-\infty}^\infty d\omega f(\omega) e^{i\omega t}. \notag
\end{align}}, where $(\pm i \omega)^\alpha$ is the principal value of the complex power function~\cite{complexpower}. The sign of the frequency determines the rotational direction in complex space, which makes their combination of particular interest, since only scaling remains: The combination
\begin{align}
    \prescript{W}{-}{\bm{D}_t^\alpha}\prescript{W}{+}{\bm{D}_t^\alpha}f(t) &= \mathcal{F}^{-1}\otherbracket{|\omega|^{2\alpha}\mathcal{F}\squarebracket{f(t);\omega}; t}
\end{align}
is known as the \textit{fractional Laplacian} $\roundbracket{-\Box_t}^{2\alpha}$~\cite{MUSINA20161667,LISCHKE2020109009}. In particular, $\alpha = 1/4$ yields the operator $\sqrt{-\Box_t}$ which is encountered among others in the context of Pseudo QED \cite{MARINO1993551,pseudoqed}. This theory emerges from a projected QED~\cite{MARINO1993551} because in 2D systems like graphene, the dynamics of the electrons is restricted to a 2D plane, whereas the photons intermediate their interactions in 3D. Fractional Laplacians were shown to be relevant to describe the quantum valley Hall effect in graphene~\cite{marino2015interaction}, as well as excitons in transition metal dichalcogenides~\cite{Marino_2018_sec,kirichenko2021influence}.

The Weyl and pseudo Weyl fractional derivatives are linear and real operators. In addition, they commute $\prescript{W}{-}{\bm{D}_t^{\vphantom{\beta}\alpha}}\prescript{W}{+}{\bm{D}_t^{\vphantom{\alpha}\beta}} = \prescript{W}{+}{\bm{D}_t^{\vphantom{\alpha}\beta}}\prescript{W}{-}{\bm{D}_t^{\vphantom{\beta}\alpha}}$ and satisfy the semigroup property, meaning that their orders add up. Some very important properties of the (pseudo) Weyl fractional derivative are that their convolution satisfies
\begin{align}
g(t)\ast \prescript{W}{\pm}{\bm{D}^\alpha_t}f(t) = \prescript{W}{\pm}{\bm{D}^\alpha_t}g(t) \ast f(t) =  \prescript{W}{\pm}{\bm{D}^\alpha_t} \squarebracket{g(t)\ast f(t)}
\label{eq:convolution_weyl}
\end{align}
and their partial integration formula reads
\begin{align}
\int_{-\infty}^{\infty}dt\; g(t)\prescript{W}{+}{\bm{D}^\alpha_t}f(t) &= \int_{-\infty}^{\infty}dt \; f(t)\prescript{W}{-}{\bm{D}^\alpha_t}g(t),
\label{eq:PI}
\end{align}
where we call the attention to the change of sign, from $\prescript{W}{+}{\bm{D}^\alpha_t}$ to $\prescript{W}{-}{\bm{D}^\alpha_t}$. The proofs of these statements are shown in the Appendix \ref{app:frac_derv}. By making use of Eq.~\eqref{eq:PI}, one can derive the Euler-Lagrange equations for Lagrangians of the form
\begin{align}
    \mathcal{L}\left( t, y(t), \left\{ \prescript{W}{+}{\bm{D}_t^{\alpha_k}}y(t), \prescript{W}{-}{\bm{D}_t^{\beta_j}}y(t) \right\}_{k,j}\right),
    \label{eq:lagragians_general}
\end{align}
where $k = 1, ...,n$, $j=1,...,m$, and all $\alpha_k$'s and $\beta_j$'s are different from each other and non-zero to avoid over-counting. In this case, the \textit{Weyl fractional Euler-Lagrange equation} reads
\begin{align}
    \frac{\partial \mathcal{L}}{\partial y} + \sum_{k = 1}^n \prescript{W}{-}{\bm{D}_t^{\alpha_k}}\frac{\partial\mathcal{L}}{\partial \prescript{W}{+}{\bm{D}_t^{\alpha_k}} y} + \sum_{j = 1}^m \prescript{W}{+}{\bm{D}_t^{\beta_j}}\frac{\partial\mathcal{L}}{\partial \prescript{W}{-}{\bm{D}_t^{\beta_j}} y} = 0.
    \label{eq:fractional_euler_lagrange_eq}
\end{align}
A detailed derivation of the above equation can be found in App.~\ref{app:frac_derv}. It is partially inspired by Ref.~\cite{AGRAWAL2002368}, which however used a Riemann-Liouville fractional derivative to a simpler version of Eq.~\eqref{eq:fractional_euler_lagrange_eq}. Note that some of the derivatives could be of integer order. This is not excluded. 

\section{Derivation of the Weyl fractional Langevin equation}
\label{sec3}

The fractional Caldeira-Leggett model is an extension of the original model, in which we modify the interaction term by introducing a Weyl fractional derivative in the coupling between the system and bath. 
The rational behind this modification is that the particle and the bath remain exactly the same, however the interaction between them changes. Ref.~\cite{ruward} showed the effect of modifications in this term. The alteration of the coupling could be attributed to several external factors, such as the geometry in which the particle or bath resides or the interaction mechanism. We  consider the Lagrangian
\begin{align}
    \mathcal{L} = \mathcal{L}_p + \mathcal{L}_{bath} + \mathcal{L}_{int} + \mathcal{L}_{CT},
    \label{eq:lag1}
\end{align}
where
\begin{align}
    \mathcal{L}_p &= \frac{1}{2}M\dot{Q}^2 - V(Q), \\
    \mathcal{L}_{bath} &= \frac{1}{2}\sum_{k = 1}^Nm_k(\dot{q_k}^2 - \omega_k^2q_k^2), \\
    \mathcal{L}_{int} &= \prescript{W}{+}{\bm{D}_t^\alpha}Q\sum_{k=1}^N C_kq_k.
\end{align}
Here, $Q$ is the coordinate, $M$ is the mass of the system, and $V(Q)$ is the potential that the system undergoes. The harmonic oscillators of the bath have coordinates $q_k$, frequency $\omega_k$, and mass $m_k$. $\alpha \in \mathbb{R}$ is the order of the fractional derivative and $\mathcal{L}_{CT}$ is a counter-term Lagrangian. It is obtained by imposing that the system of interest must be subject only to the potential $V(Q)$ and not to some renormalized potential due to the coupling with the reservoir. Therefore, we must impose that $\frac{\partial \mathcal{L}}{\partial Q} \stackrel{!}{=}-\frac{\partial V}{\partial Q}$. To do this, we first set
\begin{align}
0 &= \frac{\partial \mathcal{L}}{\partial q_j} = -m_j\omega_j^2 q_j + C_j\prescript{W}{+}{\bm{D}_t^\alpha}Q,
\end{align}
because we want the bath to be non-interacting, meaning that each particle of the bath should not feel any potential. This implies that
\begin{align}
q_j &= \frac{C_j}{m_j\omega_j^2}\prescript{W}{+}{\bm{D}_t^\alpha}Q.
\label{eq:extrje}
\end{align}
Now, we can find the minimum of $\mathcal{L}$ with respect to $Q$, which should be the minimum of the potential, such that
\begin{align}
    \frac{\partial\mathcal{L}}{\partial Q} &= -\frac{\partial V}{\partial Q} + \frac{\partial}{\partial Q}\roundbracket{\prescript{W}{+}{\bm{D}_t^\alpha} Q}\sum_{k=1}^N C_kq_k + \frac{\partial \mathcal{L}_{CT}}{\partial Q}\notag \\
    &\stackrel{!}{=}-\frac{\partial V}{\partial Q}.
    \label{eq:nvdh}
\end{align}
Upon substituting Eq.~\eqref{eq:extrje} into Eq.~\eqref{eq:nvdh} and solving for $\mathcal{L}_{CT}$, we find 
\begin{align}
\mathcal{L}_{CT} = -\frac{1}{2}\sum_{k=1}^N \frac{C_k^2}{m_k\omega_k^2}\roundbracket{\prescript{W}{+}{\bm{D}_t^\alpha} Q}^2.
\label{eq:counterfract}
\end{align}

Now that we have the full Lagrangian, we can study the equations of motion. The Euler-Lagrange equations for the bath are given by
\begin{align}
m_j\ddot{q_j} = -m_j\omega_j^2q_j + C_j \prescript{W}{+}{\bm{D}_t^\alpha} Q.
\label{eq:bath}
\end{align}
Because we want to solve equations using Fourier transforms, we need to find two solutions: a particular solution, which has no boundary conditions, and a homogeneous solution, in which we set the dependence of external functions to zero, yielding precisely the boundary conditions. Taking the Fourier transform of Eq.~\eqref{eq:bath} yields the particular solution, $q_j^p(t)$,
\begin{align}
q_j^p(\omega) &=-\frac{C_j}{m_j(\omega^2 - \omega_j^2)}(i\omega)^\alpha Q(\omega).
\label{eq:bathp}
\end{align}
In time, this can be written as
\begin{align}
    q_j^p(t) =& \frac{C_j}{m_j\omega_j^2} \left\{ \prescript{W}{+}{\bm{D}_t^\alpha} Q(t)\right.\nonumber\\
    &-\left. \frac{d}{dt} \prescript{W}{+}{\bm{D}_t^\alpha} \squarebracket{\cos (\omega_j t)\Theta(t)\ast Q(t)}\right\}.
\label{eq:bathpreal}
\end{align}
The details of this derivation can be found in the Appendix \ref{sec:bath_elminating}. The homogeneous part of the solution, $q_j^h(t)$, is given by
\begin{align}
    q_j^h(t) = A\cos(\omega_jt) + B\sin(\omega_jt),
    \label{eq:bathhomogen}
\end{align}
where $A$ and $B$ are constants determined by the initial conditions. Thus, the full solution of Eq.~\eqref{eq:bath} is given by
\begin{align}
    q_j(t) &= q_j^p(t)+q_j^h(t).
    \label{eq:bathcoord}
\end{align}

Now that we have found the solutions for the bath, we can use these in the equation of motion of the system, which can be derived from the Weyl fractional Euler-Lagrange equation. By applying Eq.~\eqref{eq:fractional_euler_lagrange_eq} to Eq.~\eqref{eq:lag1}, we obtain
\begin{align}
M\ddot{Q}(t) = &- \frac{\partial V(Q)}{\partial Q} +\sum_{k=1}^N\Bigg[ C_k  \prescript{W}{-}{\bm{D}_t^\alpha}q_k(t) \notag \\ 
&- \frac{C_k^2}{m_k\omega_k^2}\prescript{W}{-}{\bm{D}_t^\alpha}\prescript{W}{+}{\bm{D}_t^\alpha} Q(t)\Bigg].
\label{eq:above}
\end{align}
The bath coordinates can be eliminated by inserting Eqs.~\eqref{eq:bathpreal}-\eqref{eq:bathcoord} into Eq.~\eqref{eq:above}, and we eventually obtain
\begin{align}
    &M\ddot{Q}(t) = -\frac{\partial V(Q)}{\partial Q}\nonumber\\
    &-\sum_{k=1}^N \Bigg\{\frac{C_k^2}{m_k\omega_k^2}\frac{d}{dt}\prescript{W}{-}{\bm{D}_t^\alpha}\prescript{W}{+}{\bm{D}_t^\alpha}\squarebracket{\cos (\omega_kt)\Theta(t) \ast Q(t)} \notag \\
    &+C_kA\prescript{W}{-}{\bm{D}_t^\alpha}\cos(\omega_kt)+C_kB\prescript{W}{-}{\bm{D}_t^\alpha}\sin(\omega_kt)\Bigg\}.
\label{eq:intermediate}
\end{align}

Now, we continue by calculating the bath spectral function associated with the Lagrangian \eqref{eq:lag1}, given by the imaginary part of the retarded dynamical susceptibility of the bath
\begin{align}
    J(\omega)&= \text{Im}\left[-i \theta(t-t') \left\langle \left[F(t),F(t')\right]\right\rangle  \right]\nonumber\\
    &= \text{Im}\left[\chi_B(\omega) \right],
\end{align}
where $F(t)$ is the force acting on the bath by the system. The spectral function can be interpreted as describing how much of each frequency is available in the bath to interact with the system. The force coming from the interaction term in Fourier space is given by $F(\omega) = (i\omega)^\alpha \sum_{k=1}^N C_kq_k(\omega)$. Eq.~\eqref{eq:bathcoord} shows that $q_k(\omega)$ is a function of $Q(\omega)$ only via the particular solution, such that Eq.~\eqref{eq:bathp} can be directly used. The dynamical susceptibility is thus given by
\begin{align}
    \chi_B(\omega) &\equiv \frac{\partial F(\omega)}{\partial Q(\omega)}= -\sum_{k=1}^N \frac{C_k^2 (i\omega)^{2\alpha}}{m_k(\omega^2-\omega_k^2)}.
\end{align}
Writing $(i\omega)^{2\alpha} =|\omega|^{2\alpha}e^{2\alpha i \text{arg}(i\omega)} = |\omega|^{2\alpha}e^{\alpha i \pi\text{sign}(\omega)} = |\omega|^{2\alpha}\squarebracket{\cos (\alpha \pi) + i\ \text{sign}(\omega)\sin (\alpha \pi)}$ yields
\begin{align}
\chi_B(\omega) &= 
-\sum_{k=1}^N |\omega|^{2\alpha}\frac{C_k^2 }{m_k(\omega^2-\omega_k^2)} \notag \\
&\times\squarebracket{\cos (\alpha \pi) + i\ \text{sign}(\omega)\sin (\alpha \pi)}.
\label{eq:chi}
\end{align}
Interpreting $1/(\omega^2-\omega_k^2)$ as its principal value yields
\begin{align}
     \frac{1}{\omega^2 - \omega_k^2} &= \frac{\pi i}{2\omega_k}\squarebracket{\delta(\omega + \omega_k) - \delta(\omega - \omega_k)}.
     \label{eq:deltas}
\end{align}
 Substituting Eq.~\eqref{eq:deltas} into Eq.~\eqref{eq:chi} and taking the imaginary part gives
\begin{align}
    J(\omega)&=\text{Im}\squarebracket{\chi_B(\omega)}\notag \\
    &=\frac{\pi}{2}|\omega|^{2\alpha}\cos(\pi\alpha)\sum_{k=1}^N\frac{C_k^2}{m_k\omega_k}\squarebracket{\delta(\omega-\omega_k)-\delta(\omega+\omega_k)} \notag \\
    &= \frac{\pi}{2}\omega^{2\alpha}\cos(\pi\alpha)\sum_{k=1}^N\frac{C_k^2}{m_k\omega_k}\delta(\omega-\omega_k)\notag \\
    &= \omega^{2\alpha}\cos(\pi\alpha)J_{CL}(\omega),
    \label{eq:spectralfunction}
\end{align}
where in the second line we imposed that physical frequencies are positive, $\omega > 0$ and $\omega_k > 0$. In the last line, we identified the usual Caldeira-Leggett spectral function \cite{caldeira_2014}, which is given by
\begin{align}
    J_{CL}(\omega) = \frac{\pi}{2}\sum_{k=1}^N\frac{C_k^2}{m_k\omega_k}\delta(\omega-\omega_k).
    \label{eq:clspectral}
\end{align}
Eq.~\eqref{eq:spectralfunction} suggests that the Lagrangian \eqref{eq:lag1} can describe anomalous diffusion without imposing a non-Ohmic macroscopic spectral function, because the microscopic spectral function has a non-Ohmic prefactor $\omega^{2\alpha}$ that appears naturally. 
This is in agreement with investigations of spin-boson models~\cite{kehrein1996spin,anders2007equilibrium,vojta2005quantum}, where a spin-half particle is coupled to a bosonic harmonic oscillator bath. Indeed, in Ref.~\cite{vojta2005quantum}, they show that the effect of a sub-Ohmic bath can be fully characterized by a single effective interaction term, which is quadratic in the interaction. This explains why we find the modified spectral function with an extra $2\alpha$ power in the frequency. In contrast to just assuming a non-linear spectral function, however, our method provides a microscopic justification for the non-linearity from the coupling term.  In the limit of $\alpha\to 0$, we obtain the usual Caldeira Leggett spectral function.

Returning back to Eq.~\eqref{eq:intermediate}, we identify the friction force $F_{fr}(t)$ and the fluctuating force $f(t)$,
\begin{align}
&F_{fr}(t) = \sum_{k=1}^N \frac{C_k^2}{m_k\omega_k^2}\frac{d}{dt}\prescript{W}{-}{\bm{D}_t^\alpha}\prescript{W}{+}{\bm{D}_t^\alpha}\squarebracket{\cos (\omega_kt)\Theta(t) \ast Q(t)},
\label{eq:frictiondef}\\ 
&f(t) = \sum_{k=1}^NC_k\prescript{W}{-}{\bm{D}_t^\alpha}\squarebracket{A\cos(\omega_kt)+B\sin(\omega_kt)}.
\label{eq:fluctuatingdef}
\end{align}
Note that in the limit of $\alpha \to 0$, we get exactly the same friction and fluctuating forces as in the usual Caldeira-Leggett model \cite{caldeira_2014}.

Starting with the friction force, we can insert Eq.~\eqref{eq:clspectral} into Eq.~\eqref{eq:frictiondef} to obtain
\begin{align}
    F_{fr}(t) 
    &= \frac{2}{\pi}\frac{d}{dt}\prescript{W}{-}{\bm{D}_t^\alpha}\prescript{W}{+}{\bm{D}_t^\alpha}\int_{0}^t d\tau \int_{0}^{\infty} d\omega  \notag \\
    &\times \frac{J_{CL}(\omega)}{\omega}\cos[\omega(t-\tau)] Q(\tau).
\label{eq:frictemp}
\end{align}
To verify whether the above expression actually is a friction term, we need to acquire more knowledge of $J_{CL}(\omega)$. The spectral function tells us how the bath reacts to a given frequency, which means that $J_{CL}(0) = 0$ if the bath is in equilibrium. Therefore, we can perform a Taylor approximation of $J_{CL}(\omega)$ to get linear behavior for small frequencies. This is effectively the same as taking the Caldeira-Leggett spectral function to be Ohmic,
\begin{align}
\begin{split}
    J_{CL}(\omega) = \begin{cases}
    & \eta \omega \quad \text{ for } \omega < \Omega, \\
    & 0 \qquad \text{for } \omega > \Omega,
    \end{cases}
\end{split}
\end{align}
where $\Omega$ is a high-frequency cut-off ($\Omega \to \infty$) and $\eta$ is a macroscopic friction coefficient. Although the spectral function could in principle have any shape, this is the most common assumption in literature \cite{caldeira_2014, Zheng2004}. With this assumption, we see that the frequency  only remains in the argument of the cosine in Eq.~\eqref{eq:frictemp}, which is a delta function representation. Hence, we find that
\begin{align}
     F_{fr}(t) &= 2\eta\frac{d}{dt}\prescript{W}{-}{\bm{D}_t^\alpha}\prescript{W}{+}{\bm{D}_t^\alpha}\left[\int_{0}^t d\tau \delta(t-\tau) Q(\tau) \right] \notag \\
     &= \eta \prescript{W}{-}{\bm{D}_t^\alpha}\prescript{W}{+}{\bm{D}_t^\alpha}\dot{Q}(t),
     \label{eq:fractional_friction}
\end{align}
where, like in the calculation for the Caldeira-Leggett friction term, the delta function is evaluated at the boundary, which gives an extra factor of $1/2$.
Notice that Eq.~\eqref{eq:fractional_friction} reduces to a normal friction term in the limit of $\alpha \to 0$, which is indeed when our model should reduce to the usual Caldeira-Leggett model. 

Next, we turn our focus to the fluctuating force given by Eq.~\eqref{eq:fluctuatingdef}. The first step is to determine the constants $A$ and $B$ in Eq.~\eqref{eq:bathcoord}. This is done by only coupling the particle coordinate $Q(t)$ to the bath after $t=0$, such that $A$ and $B$ can simply be found by inserting $t=0$ in Eq.~\eqref{eq:bathhomogen}. We thus find
\begin{align}
    A = q_j(0), \quad
    B =\dot{q_j}(0)/\omega_j.
\end{align}
Inserting this into Eq.~\eqref{eq:frictiondef} yields
\begin{align}
    f(t) = \sum_{k=1}^NC_k\prescript{W}{-}{\bm{D}_t^\alpha}\squarebracket{q_k(0)\cos(\omega_kt)+\frac{\dot{q_k}(0)}{\omega_k}\sin(\omega_kt)}.
    \label{eq:fluctating_force_fin}
\end{align}
Eq.~\eqref{eq:fluctating_force_fin} is the same as the fluctuating force in the usual Caldeira-Leggett model up to $\prescript{W}{-}{\bm{D}_t^\alpha}$. Since this operator is linear we ignore it during the calculation such that we can proceed analogously to the calculation done for the Caldeira-Leggett model~\cite{timeglass}. We then find
\begin{align}
    &\langle f(t) \rangle = 0, \label{eq:average_fluc}\\
    &\langle f(t)f(t') \rangle = 2\eta k_B T \prescript{W}{-}{\bm{D}_t^\alpha}\prescript{W}{-}{\bm{D}_{t'}^\alpha}\delta(t-t').
    \label{eq:correlator_fluc}
\end{align}
 We found that Eq.~\eqref{eq:correlator_fluc} actually describes colored noise. Indeed, using the definition of the pseudo Weyl fractional derivative acting on the Fourier representation of the delta function, we obtain
\begin{align}
    \prescript{W}{-}{\bm{D}_t^\alpha}\prescript{W}{-}{\bm{D}_{t'}^\alpha}\delta(t-t')&= \frac{\sin(\pi\alpha)\Gamma(2\alpha + 1)}{\pi}\roundbracket{t-t'}^{-2\alpha -1}.
    \label{eq:colored_noise_weyl}
\end{align}
This representation of colored noise is only valid for $(2\alpha + 1) > 0$.

Finally, substituting Eqs.~\eqref{eq:fractional_friction} and \eqref{eq:fluctating_force_fin} into Eq.~\eqref{eq:intermediate} yields
\begin{align}
    M\ddot{Q}(t) + \frac{\partial V(Q)}{\partial Q} + \eta \prescript{W}{-}{\bm{D}_t^\alpha}\prescript{W}{+}{\bm{D}_t^\alpha}\dot{Q}(t) = f(t),
    \label{eq:fractlang1}
\end{align}
with $\langle f(t)f(t') \rangle = 2\eta k_B T \prescript{W}{-}{\bm{D}_t^\alpha}\prescript{W}{-}{\bm{D}_{t'}^\alpha}\delta(t-t')$ and $\langle f(t) \rangle = 0$. Note that just by changing the coupling to the bath, we obtained a modified version of the Langevin equation that we call the \textit{Weyl fractional Langevin equation}. There was no need to assume a non-Ohmic bath explicitly. This is the first main result of of this paper.


\section{Solving the Weyl fractional Langevin equation for the free case}
\label{sec4}

When the particle is not subject to any external potential, Eq.~\eqref{eq:fractlang1} reduces to
\begin{align}
    M\ddot{Q}(t) + \eta \prescript{W}{-}{\bm{D}_t^\alpha}\prescript{W}{+}{\bm{D}_t^\alpha}\dot{Q}(t) = f(t).
    \label{eq:fractlang_free}
\end{align}
Taking the Fourier transform of Eq.~\eqref{eq:fractlang_free}, rearranging and taking the inverse Fourier transform, leads to the particular solution
\begin{align}
   Q^p(t) &= \mathcal{F}^{-1}\squarebracket{\frac{f(\omega)}{-M\omega^2 + \eta i \omega|\omega|^{2\alpha}}; t} \notag \\
   &= \frac{1}{2\pi}f(t)\ast \mathcal{F}^{-1}\roundbracket{\frac{1}{-M\omega^2 + \eta i \omega|\omega|^{2\alpha}}; t}.
   \label{eq:part_solut}
\end{align}
The homogeneous solution of Eq.~\eqref{eq:fractlang_free} is obtained by solving $\omega\roundbracket{-M\omega + \eta i|\omega|^{2\alpha}} = 0,$ thus yielding $\omega = 0 \text{ or } \omega = i \roundbracket{\eta/M}^{1/(1-2\alpha)}.$ Hence, the homogeneous solution reads
\begin{align}
    Q^h(t) &= \mathcal{F}^{-1}\otherbracket{A\delta(\omega) + B\delta\squarebracket{\omega - i \roundbracket{\frac{\eta}{M}}^{\frac{1}{1-2\alpha}}}} \notag \\
    &= C + D \text{exp}\squarebracket{-t\roundbracket{\frac{\eta}{M}}^{\frac{1}{1-2\alpha}}}\notag \\
    &= C + D \text{exp}\roundbracket{-\frac{t}{t_s}},
    \label{eq:homogeneoussolution}
\end{align}
where $C$ and $D$ are constants that will be determined later and $t_s = (M/\eta)^{1/(1-2\alpha)}$ is a characteristic time-scale of our system. Combining Eqs.~\eqref{eq:part_solut}~and~\eqref{eq:homogeneoussolution}, we obtain
\begin{align}
    Q(t) &= Q^p(t) + Q^h(t) \notag \\
    &= \frac{1}{2\pi}f(t)\ast \mathcal{F}^{-1}\roundbracket{\frac{1}{-M\omega^2 + \eta i \omega|\omega|^{2\alpha}}; t} \notag \\
    &+ C + D \text{exp}\roundbracket{-\frac{t}{t_s}}.
    \label{eq:fractional_full_solution}
\end{align}
To determine the constants $C$ and $D$, we proceed in the same way as for the fluctuating force. We consider that the particle is only coupled to the bath for $t>0$, such that we can take $f(0) = 0$. This implies that $C$ and $D$ are determined solely by the homogeneous part in Eq.~\eqref{eq:fractional_full_solution}, which leads to
\begin{align}
\begin{split}
    \begin{cases}
    C = Q(0) + \dot{Q}(0)t_s, \\
    D = -\dot{Q}(0)t_s.
    \end{cases}
    \end{split}
\end{align}
Since our system is translation invariant, we can chose for simplicity $Q(0) = 0$. Eq.~\eqref{eq:fractional_full_solution} then becomes
\begin{align}
    Q(t) &= \frac{f(t)}{2\pi}\ast \mathcal{F}^{-1}\roundbracket{\frac{1}{-M\omega^2 + \eta i \omega|\omega|^{2\alpha}}; t} \notag \\
    &+\dot{Q}(0)t_s\squarebracket{1-\text{exp}\roundbracket{-\frac{t}{t_s}}}.
\end{align}
Next, we can use that $\langle f(t) \dot{Q}(0) \rangle = \langle f(t) \rangle = 0$, $\langle \dot{Q}(0)^2 \rangle= k_BT/M$ and $\langle f(t)f(t') \rangle = 2\eta k_B T \prescript{W}{-}{\bm{D}_t^\alpha}\prescript{W}{-}{\bm{D}_{t'}^\alpha}\delta(t-t')$, to calculate the MSD for the above solution. After partial integration, we find
\begin{align}
 \left\langle Q(t)^2 \right\rangle
    &= \frac{\eta k_B T}{2\pi^2}\int_0^t d\tau \otherbracket{\mathcal{F}^{-1}\squarebracket{\frac{\roundbracket{i\omega}^\alpha}{-M\omega^2 + \eta i \omega|\omega|^{2\alpha}}; \tau}}^2 \notag \\
    &+ \frac{k_B T}{M}t_s^2\squarebracket{1-\text{exp}\roundbracket{-\frac{t}{t_s}}}^2.
    \label{eq:solution_frac_lang}
\end{align}
This is the second main result of this paper. Next, we discuss its physical implications.
\begin{figure}[t]
\centering
\includegraphics[keepaspectratio=true,width=\columnwidth]{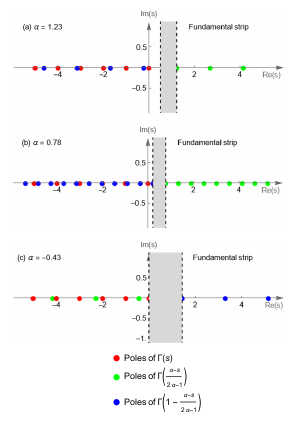}
    \caption{The three different fundamental strips of a Mellin transform involving the shown gamma functions in the legend. (a) the first fundamental strip, (b) the second and (c) the third. Notice that green poles are on the right of the fundamental strip in (a) and (b) but on the left and (c).}
    \label{fig:poles_structure}
\end{figure}

\section{Asymptotic expansions of the mean squared displacement}
\label{sec5}

To determine the asymptotic behavior of Eq.~\eqref{eq:solution_frac_lang}, we need to perform the asymptotic expansions of the inverse Fourier integral inside that equation. To achieve this aim, we first simplify the integral by using the property of the inverse Fourier of a product. Then, we take the Mellin transform of the remaining integral. Next, we write the expression in a way that is compatible with the definition of the Fox H-function. After that, we take the inverse Mellin transform and identify the specific Fox H-function that solves the problem. Finally, we use known asymptotic expansions of the Fox H-function to obtain the asymptotic behavior of Eq.~\eqref{eq:solution_frac_lang}. This procedure was inspired by Ref.~\cite{stappenplan}. Since the reader might not be familiar with the Mellin transforms and the Fox H-function, we provide a short overview of their properties in Apps.~\ref{app:mellin} and \ref{app:foxh}, respectively. More specifically, we start by analysing
\begin{align}
    I(t) \equiv \mathcal{F}^{-1}\squarebracket{\frac{\roundbracket{i\omega}^\alpha}{-M\omega^2 + \eta i \omega|\omega|^{2\alpha}}; t},
    \label{eq:fourier_int_def}
\end{align}
which can be simplified to
\begin{align}
    I(t) &= i \Theta(t) \ast \mathcal{F}^{-1}\squarebracket{\frac{\roundbracket{i\omega}^\alpha}{-M\omega + \eta i|\omega|^{2\alpha}}; t} \notag \\
    &\equiv i \Theta(t) \ast I'(t).
    \label{eq:integeral_t}
\end{align}
Now, we focus solely on $I'(t)$. Because of convergence, we will take $\alpha \neq 0, 1/2$. The Mellin transform of $I'(t)$ is given by
\begin{align}
    F(s) &= -\frac{2\pi i}{M|2\alpha-1|}t_s^{s-\alpha} \times \notag \\
    &\times \frac{\Gamma(s)\Gamma\roundbracket{1- \frac{\alpha-1}{2\alpha -1} - \frac{s}{2\alpha - 1}}\Gamma\roundbracket{\frac{\alpha-1}{2\alpha -1} + \frac{s}{2\alpha -1}}
    }{\Gamma\roundbracket{\frac{\alpha^2}{2\alpha - 1} + \frac{s}{2}}\Gamma\roundbracket{1-\frac{\alpha^2}{2\alpha - 1} - \frac{s}{2}}}.
    \label{eq:before_split_mellin}
\end{align}
Details of this derivation can be found in App.~\ref{app:asymptotic_msd}. The pole structure of Eq.~\eqref{eq:before_split_mellin} depends on $\alpha$, which has consequences when identifying a Fox H-function. This results in a region of convergence of the Mellin transform, known as the fundamental strip, that also depends on $\alpha$. The behavior of the different gamma functions can be observed in Fig.~\ref{fig:poles_structure}. Therefore, we separate the analysis in $3$ distinct cases: $\alpha > 1$, $ 1/2 < \alpha < 1$ and $\alpha < 1/2$, which we will refer to as the first, second and third fundamental strips, respectively. Here, we present the calculation for the first fundamental strip. The calculations for the second and third fundamental strips are similar and can be found in App.~\ref{app:asymptotic_msd}.

In the first fundamental strip, for $\alpha > 1$, we rewrite Eq.~\eqref{eq:before_split_mellin} in such a way that after taking the inverse Mellin transform of that equation, we can identify the Fox H-function given by
\wideeq{I'(t) = -\frac{2\pi i}{M|2\alpha-1|}t_s^{-\alpha}
 H^{21}_{23} \left[ \frac{t}{t_s}\ \middle| \begin{array}{ccc}
  & \left( \frac{\alpha-1}{2\alpha -1}, \frac{1}{2\alpha - 1}\right), & \left( \frac{\alpha^2}{2\alpha - 1}, \frac{1}{2}\right)\\
  (0, 1) & \left( \frac{\alpha - 1}{2\alpha - 1}, \frac{1}{2\alpha - 1} \right), & \left( \frac{\alpha^2}{2\alpha -1}, \frac{1}{2}\right) 
\end{array} \right],
\label{eq:foxH_final}}
where we again identified the characteristic timescale of the system, $t_s$.
Using the asymptotic expansion of the Fox H-function developed in App. \ref{app:foxh}, we find
\begin{equation}
    I'(t)  \sim \begin{cases}
&1 \qquad \: \text{ when } t \to 0, \\
&t^{-\alpha} \quad \ \: \text{when } t \to \infty.
\end{cases}
\label{eq:assymp_first_strip}
\end{equation}
Substituting Eq.~\eqref{eq:assymp_first_strip} into Eq.~\eqref{eq:integeral_t} gives
\begin{align}
    I(t) \sim \begin{cases}
&t \qquad \: \ \ \  \text{ when } t \to 0, \\
&t^{-\alpha + 1} \quad \: \: \text{when } t \to \infty.
\label{eq:intergral_expansion_first}
\end{cases}
\end{align}
For $t\to 0$, we Taylor expand the exponential inside Eq.~\eqref{eq:solution_frac_lang} up to linear order and discard it for $t\to \infty$. Substituting Eq.~\eqref{eq:intergral_expansion_first} into Eq.~\eqref{eq:solution_frac_lang} together with the approximations of the exponential function, as explained before, results in 
\begin{align}
    \left\langle Q(t)^2 \right\rangle \sim \begin{cases}
&t^{2} \quad \quad \ \ \: \:  \text{ when } t \to 0, \\
&t^{-2\alpha + 3}  \quad \text{ when } t \to \infty,
\end{cases}
\label{eq:msd_first_fundamental_strip}
\end{align}
for $\alpha > 1$. Combining this with the results from the second and third fundamental strips derived in App. \ref{app:asymptotic_msd}, we find that the short-time behavior of the MSD is given by
\begin{align}
    \left\langle Q(t)^2 \right\rangle \stackrel{t \to 0}{\sim} \begin{cases}
        &t^2 \quad \ \ \: \: \ \text{ for } \alpha > \frac{-1 - \sqrt{5}}{4}, \alpha \notin \mathcal{P},\\
        &t^{2\alpha + 3} \quad \: \text{for } \alpha < \frac{-1-\sqrt{5}}{4}, \alpha \notin \mathcal{P},
        \end{cases}
    \label{eq:short_time_msd}
    \end{align}
with $\mathcal{P} = \big\{(-3 - \sqrt{21})/4, (-1 - \sqrt{3})/2, -1/2, 0,(1 + \sqrt{3})/2,(-3 + \sqrt{21})/4, 1/2, 1 \big\}$. On the other hand, the long-time limit yields
\begin{align}
    \left\langle Q(t)^2 \right\rangle \stackrel{t \to \infty}{\sim} \begin{cases}
&t^{-2\alpha + 3} \quad \: \text{ for }  \alpha > 1 ,\alpha \notin \mathcal{P},\\
&t^{2\alpha + 1}\quad  \ \: \ \text{ for } \alpha < 1, \alpha \notin \mathcal{P}.
\end{cases}
\label{eq:long_time_msd}
\end{align}
This is the third main result of this paper. Next, we discuss its implications and compare it with results in the literature.

\section{Discussion and Conclusion} 
\label{sec6}
We have introduced the Weyl fractional derivative which, due to its properties with the boundary and Fourier transforms, can be directly used inside the Lagrangian. We did this by imposing a fractional derivative in the coupling term between the system and bath, which we called a fractional Caldeira-Leggett model.  Doing so required the introduction of the Weyl fractional Euler-Lagrange equation to derive the equations of motion, which we solved for the harmonic oscillators using Fourier transforms. We noted that the fractional bath spectral function could be written in terms of a fractional power prefactor and the original Caldeira-Leggett spectral function. Inserting the commonly used Ohmic assumption for the Caldeira-Leggett spectral function, we found an effective spectral function which has a power law of order $2\alpha +1$. This led to a Weyl fractional Langevin equation with colored noise. We provided the analytical solutions for several quantities in terms of an inverse Fourier transform, which we could interpret as a Fox H-function with the help of Mellin transforms. Finally, the asymptotic limits of these solutions were calculated for a wide range of values. For short times, we found both ballistic and non-ballistic regimes. This means that, in some cases, friction can dominate the short-time dynamics. For long times we found everything ranging from sub-, normal-, and super-diffusion, as well as saturation, which could hint to a relation with glassy physics. Overall, we found a very broad range of anomalous diffusion.

\begin{figure*}[tb]
    \includegraphics[keepaspectratio=true,scale=0.7]{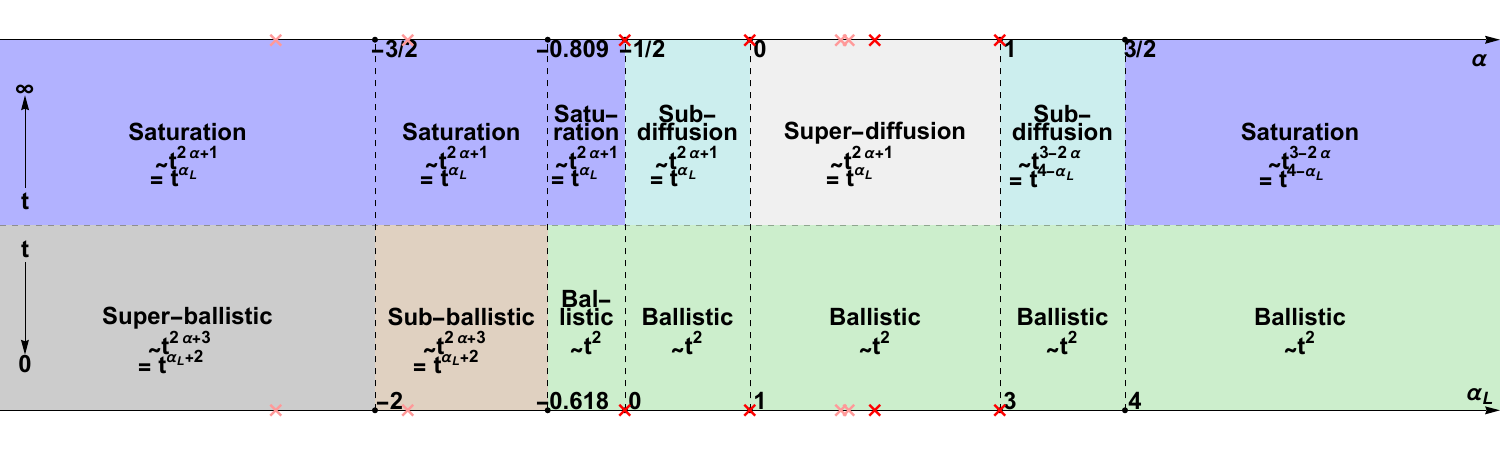}
    \caption{The behavior of the MSD for small and large times are shown in terms of the used parameter in this paper, $\alpha$, but also in terms of the total friction force order, $\alpha_L$, which are linked by $\alpha_L = 2\alpha + 1$. Some specific values of $\alpha$, and in turn $\alpha_L$, should be ignored in this plot, namely $\alpha \in \big\{(-3 \pm \sqrt{21})/4, (-1 \pm \sqrt{3})/2, \pm1/2, 0, 1 \big\}$. The light red crosses indicate the values excluded for simplicity while the red crosses indicate the values for which our calculation did not apply.}
    \label{fig:findings}
\end{figure*}

We are now in a position to compare our results, given by Eqs.~\eqref{eq:short_time_msd}~and~\eqref{eq:long_time_msd}, with the results of Ref.~\cite{e_lutz}, in which, to the best of our knowledge, a fractional Langevin equation was investigated for the first time. First, we note that we use the Weyl fractional derivative, while Ref.~\cite{e_lutz} used the Caputo fractional derivative. It is known that the Caputo fractional derivative and Weyl fractional derivatives lead to different results in a lot of cases. One such example is when the fractional derivative of the exponential function is calculated (see App.~\ref{app:frac_derv}). This makes it very interesting to compare the results. The friction force in Eq.~\eqref{eq:fractlang_free} is of order $2\alpha + 1$, which corresponds to the derivative order from Ref.~\cite{e_lutz} (which we will call $\alpha_L$). Therefore, to compare the results, we must make the substitution $\alpha = (\alpha_L-1)/2$ in Eq.~\eqref{eq:long_time_msd}. Second, we note that Ref.~\cite{e_lutz} only considered the two cases, $0<\alpha_L<1$ and $1<\alpha_L<2$. For those restrictions, our result from Eq.~\eqref{eq:long_time_msd} becomes
\begin{align}
    \left\langle Q(t)^2 \right\rangle \stackrel{t \to \infty}{\sim} t^{\alpha_L},
\label{eq:long_time_msd2}
\end{align}
which is exactly what was found in Ref.~\cite{e_lutz}. It is remarkable that the same result is obtained for a Caputo derivative instead of Weyl, and derived using completely different techniques. Moreover, we generalized the result of the Caputo derivative, which in Ref.~\cite{e_lutz} was obtained for $0<\alpha_L<2$ (corresponding to $-1/2 <\alpha<1/2$), to the entire real line, except for $\alpha=0$, $\alpha=\pm 1/2, \alpha = 1$ (corresponding to $\alpha_L=0,1,2,3$, which are well-known integer derivative cases) and $\alpha = (-3 \pm \sqrt{21})/4$, $\alpha = (-1 \pm \sqrt{3})/2$, which are cases for which a logarithm would appear because of integration. We can thus conclude that our model, given by Eq.~\eqref{eq:lag1}, describes sub-diffusion for $-1/2< \alpha < 0$, and super-diffusion for $0 < \alpha< 1$, while $1<\alpha<3/2$ yields sub-diffusion. Remarkably, when $\alpha >3/2$ and when $\alpha<-1/2$ we find a negative MSD exponent, which typically indicates saturation in the MSD~\cite{timeglass}, possibly describing glassy states. The short-time behavior of the MSD exhibits the usual ballistic ($\sim t^2$) behavior when $\alpha > -(1+\sqrt{5})/4 \approx -0.809$. However, when $-3/2<\alpha < -(1+\sqrt{5})/4$ we find sub-ballistic behavior, meaning that the MSD  goes as $t^p$, where $0<p<2$, when $t \to 0$. In particular, we have that $0<p< (5-\sqrt{5})/2 \approx 1.382$. Finally, for $\alpha < -3/2$, we find super-ballistic behavior of the MSD for $t\to 0$, which in our words means that the short time behavior goes as a negative exponent. The different behavior of the MSD for short and long times in terms of the parameter of our model, $\alpha$, and the total order of the friction force, $\alpha_L$, is summarized in Fig.~\ref{fig:findings}.

When we calculated the spectral function, we were able to identify the regular Caldeira-Leggett spectral function inside it, but there is some delicate interpretation that needs to be done here. The spectral function describes the susceptibility of the bath to an external system in the frequency domain. However, as it turns out, this is not just a property of the bath, but also depends on the way that you couple to it, which is almost always linear. This is done because it is typically the lowest-order contribution, and it gives excellent options to proceed mathematically, with e.g. completing the square. Since baths are typically thought to be Ohmic (linear in frequency) in this fashion, we believe that the most likely assumption should be to keep this part identical, and thus retain an Ohmic $J_{CL}(\omega)\sim \omega$. In turn, this means that the fractional coupling spectral function goes as  $J(\omega) \sim \omega^{2\alpha+1}$, which resulted in a Weyl fractional Langevin equation. The results of the latter remarkably match with Ref.~\cite{e_lutz}, despite a different definition being used there, and even expand the domain of available parameters. Previously, this equation has only been derived using non-Ohmic baths, but this derivation shows that it can also be achieved for Ohmic baths, as long as the coupling term is changed accordingly. Hence, understanding the coupling mechanism is as important as understanding the bath properties in order to describe (anomalous) quantum diffusion.



As we have demonstrated, the Weyl fractional Langevin equation can describe anomalous diffusion for many different exponents. In experiments of quantum transport on fractals~\cite{exp1}, theory of diffusion on confining geometries~\cite{jeon2010fractional}, and random walks~\cite{METZLER20001,kobelev2000fractional}, several connections have been made to link either the order of the derivative or the diffusion exponent itself to a \textit{fractal dimension}. Fractals can exhibit a dimension that is non-integer, such as the Sierpiński gasket which has a dimension of $d_f\approx 1.58$. Ref.~\cite{exp1} has shown that the MSD exhibits a regime where the exponent is equal to $d_f$. Although the mechanism for this is still not fully understood, this could be explained by an effect of the fractal dimension on the interaction between the underlying material and the quantum particle. If we assume that $\alpha= (d_f -1)/2$, then the Weyl fractional Langevin equation reproduces this result.

Moving forward, an important aspect that requires further investigation is the realization of a fractional derivative coupling as an effective theory. One promising approach is to couple the bath to another bath and subsequently integrate out the intermediate bath degrees of freedom to obtain an effective bath with the desired fractional derivative coupling, or perhaps by some interaction within the bath that could add frequency dependence. Another potential method that could result in Eq.~\eqref{eq:lag1} is inspired by the projection scheme used in Refs.~\cite{MARINO1993551,pseudoqed} and the fractal delta function used in Ref.~\cite{fractaldelta}. In this approach, one would project the initial theory down to a fractal, which could result in a fractional coupling. Additionally, applying fractal statistics to the bath could lead to the fractional derivative coupling. Furthermore, it is also interesting to explore quantization for the Weyl fractional derivative further than has already been done in the literature \cite{qm0982, PhysRevEfracpath,weylquan}. To do this, one would first need to develop the Hamiltonian formalism for these fractional derivatives, which can be inspired by Ref.~\cite{BALEANU2008199}. After that, one chooses either for canonical quantization~\cite{qm0982}, path integral quantization~\cite{PhysRevEfracpath} or the less known Weyl-Wigner quantization~\cite{weylquan}. The last is in our opinion the most natural one to use for the Weyl fractional derivative and therefore the most promising method. Finally, we believe that it would be interesting to see if previous works \cite{kehrein1996spin,verstraten2022fractional,nalbach2010ultraslow}, that assumes a non-Ohmic bath, could be reproduced by the approach used in this paper, as it would introduce a new way to justify the models microscopically.


\begin{acknowledgments}
We thank Lars Fritz and Rodrigo Arouca for fruitful discussions. 
This work was supported by the
Netherlands Organization for Scientific Research (NWO,
Grant No. 680.92.18.05, C.M.S. and R.C.V.).
\end{acknowledgments}


\appendix

\section{Fractional derivative properties}
\label{app:frac_derv}

As shortly mentioned in the introduction, there exist many fractional derivatives. One that is often used in the literature \cite{timeglass, e_lutz} is the Caputo fractional derivative, defined as
\begin{align}
\prescript{C}{}{\bm{D}_t^s} f(t) = \frac{1}{\Gamma(n-s)}\int_0^t d\tau \roundbracket{t-\tau}^{n-s-1}f^{(n)}(\tau),
\end{align}
where $n \in \mathbb{N}$ such that $n-1 \leq s < n$. However, the Caputo derivative is not compatible with Fourier transforms, and for this reason we use the Weyl fractional derivative, which is defined in Eq. \eqref{eq:def_weyl} in this paper. To illustrate the difference between the Caputo and Weyl fractional derivative, we shortly discuss what happens to the exponential function when applying both of these fractional derivatives to it. One of the defining properties of the exponential function is that it is an eigenfunction of the derivative, meaning that the derivative of the exponential function is itself an exponential function up to some scaling. When generalizing integer order derivatives, this might be one of the properties that one wants to keep. The Weyl fractional derivative is one of those generalizations that satisfies this property, namely
\begin{align}
    \prescript{W}{+}{\bm{D}_t^\alpha} e^{at} &= \mathcal{F}^{-1}\otherbracket{(i\omega)^\alpha\mathcal{F}\squarebracket{e^{at};\omega}; t} \notag \\
    &= \mathcal{F}^{-1}\otherbracket{(i\omega)^\alpha\delta(\omega+ia) ;t} \notag \\
    &=a^\alpha e^{at}.
\end{align}
This is in stark contrast with what happens when one applies the Caputo fractional derivative to the exponential function. In that case, we obtain
\begin{align}
    \prescript{C}{}{\bm{D}_t^s} e^{at} &= \sum_{i = 0}^\infty \frac{\Gamma(n+1)}{\Gamma(n-s+1)}t^{n-\alpha}.
\end{align}
Thus, the exponential function is clearly not an eigenfunction of the Caputo fractional derivative. This is one of the many differences between the Caputo and Weyl fractional derivatives.

In what follows, we present mathematical proofs of certain properties of the Weyl fractional derivative. First, we note that the (pseudo) Weyl fractional derivative is a linear operator, which directly follows from the definition and linearity of integrals. Next, we show that these operators are indeed real operators, meaning that
\begin{flalign}
\left[\prescript{W}{\pm}{\bm{D}_t^\alpha}f(t)\right]^* = \prescript{W}{\pm}{\bm{D}_t^\alpha}f(t).
\end{flalign}
It indeed holds that 
\begin{align}
    \left[\prescript{W}{\pm}{\bm{D}_t^\alpha}f(t)\right]^* &= \left[\int_{-\infty}^\infty d\omega e^{i\omega t}(\pm i \omega)^\alpha \int_{-\infty}^{\infty} \frac{d\tau}{2\pi} e^{-i\omega \tau} f(\tau) \right]^* \notag \\
    &= \int_{-\infty}^\infty d\omega e^{-i\omega t}(\mp i \omega)^\alpha \int_{-\infty}^{\infty} \frac{d\tau}{2\pi} e^{i\omega \tau} f(\tau)\notag \\
    &= \int_{-\infty}^\infty d\omega e^{i\omega t}(\pm i \omega)^\alpha \int_{-\infty}^{\infty} \frac{d\tau}{2\pi} e^{-i\omega \tau} f(\tau) \notag \\
    &= \prescript{W}{\pm}{\bm{D}_t^\alpha}f(t).
\end{align}
In addition,
\begin{align}
    \prescript{W}{-}{\bm{D}_t^{\vphantom{\beta}\alpha}}\prescript{W}{+}{\bm{D}_t^{\vphantom{\alpha}\beta}}f(t) &= \mathcal{F}^{-1}\otherbracket{(-i\omega)^\alpha\mathcal{F}\squarebracket{\prescript{W}{+}{\bm{D}^\alpha_t}f(t);\omega}; t} \notag \\
&= \mathcal{F}^{-1}\otherbracket{(-i\omega)^\alpha(i\omega)^\beta \mathcal{F}\squarebracket{f(t);\omega}; t}\notag \\
&= \mathcal{F}^{-1}\otherbracket{(i\omega)^\beta \mathcal{F}\squarebracket{\prescript{W}{-}{\bm{D}^\alpha_t}f(t);\omega}; t} \notag \\
&= \prescript{W}{+}{\bm{D}_t^{\vphantom{\alpha}\beta}}\prescript{W}{-}{\bm{D}_t^{\vphantom{\beta}\alpha}}f(t),
\end{align}
which means that the pseudo and the Weyl fractional derivative commute. They also satisfy the semi-group property, meaning that their orders add up. This is proved by 
\begin{align}
    \prescript{W}{\pm}{\bm{D}_t^{\vphantom{\alpha}\beta}}\prescript{W}{\pm}{\bm{D}_t^{\vphantom{\beta}\alpha}}f(t) &= \mathcal{F}^{-1}\otherbracket{(\pm i\omega)^\alpha(\pm i\omega)^\beta \mathcal{F}\squarebracket{f(t);\omega}; t}\notag \\
    &= \mathcal{F}^{-1}\otherbracket{(\pm i\omega)^{\alpha + \beta}\mathcal{F}\squarebracket{f(t);\omega}; t}\notag \\
    &= \prescript{W}{\pm}{\bm{D}^{\alpha+\beta}_t}f(t).
\end{align}
Next, we will proof Eq. \eqref{eq:convolution_weyl}. The Fourier transform of a convolution, where one of the terms is a Weyl fractional derivative of a function $f$, can be rewritten as 
\begin{align}
    \mathcal{F}\squarebracket{g(t)\ast \prescript{W}{\pm}{\bm{D}^\alpha_t}f(t) ;\omega} &= \mathcal{F}\squarebracket{g(t); \omega}\mathcal{F}\squarebracket{\prescript{W}{\pm}{\bm{D}^\alpha_t}f(t); \omega} \notag \\
     &= \mathcal{F}\squarebracket{g(t); \omega}(\pm i\omega)^\alpha\mathcal{F}\squarebracket{f(t); \omega} \notag \\
     &= \mathcal{F}\squarebracket{\prescript{W}{\pm}{\bm{D}^\alpha_t}g(t); \omega}\mathcal{F}\squarebracket{f(t); \omega} \notag \\
     &= \mathcal{F}\squarebracket{\prescript{W}{\pm}{\bm{D}^\alpha_t}g(t)\ast f(t) ;\omega}.
\end{align}
Furthermore, we know that
 \begin{align}
     \mathcal{F}\otherbracket{\prescript{W}{\pm}{\bm{D}^\alpha_t}\squarebracket{g(t)\ast f(t) };\omega} &= \mathcal{F}\squarebracket{g(t); \omega}(\pm i\omega)^\alpha\mathcal{F}\squarebracket{f(t); \omega}.
 \end{align}
 Hence, we conclude, by uniqueness of the Fourier transform, that 
 \begin{align}
g(t)\ast \prescript{W}{\pm}{\bm{D}^\alpha_t}f(t) = \prescript{W}{\pm}{\bm{D}^\alpha_t}g(t) \ast f(t) =  \prescript{W}{\pm}{\bm{D}^\alpha_t} \squarebracket{g(t)\ast f(t)}.
\label{eq:property_weyl_convolution}
\end{align}
Next, we will proof the partial integration formula, namely Eq. \eqref{eq:PI}. By definition
\begin{align}
    \int_{-\infty}^{\infty}dtg(t)\prescript{W}{+}{\bm{D}^\alpha_t}f(t) &=  \int_{-\infty}^{\infty}dtg(t)\int_{-\infty}^{\infty}d\omega e^{i\omega t}(i\omega)^\alpha\times \notag \\
    &\times \frac{1}{2\pi}\int_{-\infty}^{\infty} d\tau  e^{-i\omega \tau}f(\tau) \notag \\
    &= \int_{-\infty}^{\infty} d\tau f(\tau)\int_{-\infty}^{\infty} d\omega e^{i\omega \tau} (-i\omega)^\alpha \times \notag \\
    &\times \frac{1}{2\pi}\int_{-\infty}^{\infty} dt e^{-i\omega t} g(t) \notag \\
    &= \int_{-\infty}^{\infty}dtf(t)\prescript{W}{-}{\bm{D}^\alpha_t}g(t),
\end{align}
where, in the second line, we made the substitution $\omega \to -\omega$ and in the last line we renamed $t \longleftrightarrow \tau $. Finally, we prove that the equations of motion for Lagrangians of the form given in Eq. \eqref{eq:lagragians_general}, are given by Eq. \eqref{eq:fractional_euler_lagrange_eq}. We start by taking a variation of the action, which results in
\begin{align}
    \delta S[y] = \int_{-\infty}^\infty dt \Bigg( &\frac{\partial \mathcal{L}}{\partial y}\delta y + \sum_{k=1}^n\frac{\partial \mathcal{L}}{\partial \prescript{W}{+}{\bm{D}_t^{\alpha_k}}y} \delta\prescript{W}{+}{\bm{D}_t^{\alpha_k}}y \notag \\
    &+ \sum_{j=1}^m\frac{\partial \mathcal{L}}{\partial \prescript{W}{-}{\bm{D}_t^{\beta_j}}y} \delta\prescript{W}{-}{\bm{D}_t^{\beta_j}}y\Bigg).
\end{align}
Swapping the order of the variation and the fractional derivatives and applying the partial integration formula yields
\begin{align}
    \delta S[y] =\int_{-\infty}^\infty dt \Bigg(&\frac{\partial \mathcal{L}}{\partial y} + \sum_{k = 1}^n \prescript{W}{-}{\bm{D}_t^{\alpha_k}}\frac{\partial\mathcal{L}}{\partial \prescript{W}{+}{\bm{D}_t^{\alpha_k}} y} \notag \\
    &+ \sum_{j = 1}^m \prescript{W}{+}{\bm{D}_t^{\beta_j}}\frac{\partial\mathcal{L}}{\partial \prescript{W}{-}{\bm{D}_t^{\beta_j}} y}\Bigg)\delta y.
    \label{eq:proof_euler_frac}
\end{align}
Because Eq.~\eqref{eq:proof_euler_frac} holds for all $\delta y$, setting that equation equal to zero implies
\begin{align}
    \frac{\partial \mathcal{L}}{\partial y} + \sum_{k = 1}^n \prescript{W}{-}{\bm{D}_t^{\alpha_k}}\frac{\partial\mathcal{L}}{\partial \prescript{W}{+}{\bm{D}_t^{\alpha_k}} y} + \sum_{j = 1}^m \prescript{W}{+}{\bm{D}_t^{\beta_j}}\frac{\partial\mathcal{L}}{\partial \prescript{W}{-}{\bm{D}_t^{\beta_j}} y} = 0. 
\end{align}


\section{Eliminating the bath}
\label{sec:bath_elminating}

\begin{figure}[ht]
\includegraphics[keepaspectratio=true,width=\columnwidth]{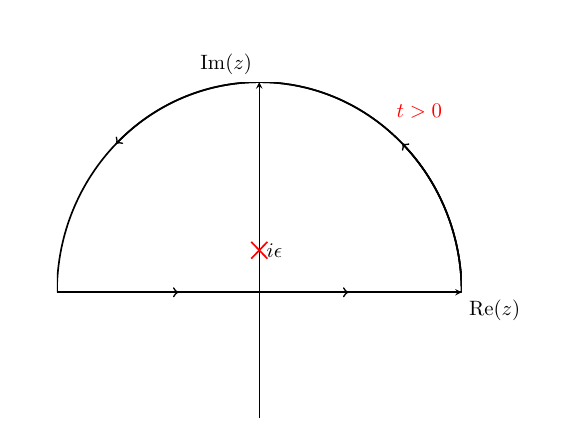}
\caption{The pole structure of the function $1/(z - i \epsilon)$ is depicted in the above figure. After applying complex analysis and taking $\epsilon \to 0$, one can give a finite value to a divergent integral over $1/z$, which we call the positive $i\epsilon$-interpretation. Note that the shown contour is only chosen for $t > 0$.}
    \label{fig:contour_first}
\end{figure}

In this appendix, we show how to pass from Eq.~\eqref{eq:bathp} to Eq.~\eqref{eq:bathpreal}, since it is highly non-trivial. Taking the inverse Fourier transform of Eq.~\eqref{eq:bathp} yields
\begin{widetext}
    \begin{align}
q_j^p(t)&= -\mathcal{F}^{-1}\left[\frac{C_j}{m_j(\omega^2 - \omega_j^2)}(i\omega)^\alpha Q(\omega); t\right] \notag \\
&=\mathcal{F}^{-1}\squarebracket{\frac{(i\omega)^\alpha C_j}{m_j\omega_j^2}\roundbracket{1-\frac{\omega^2}{\omega^2-\omega_j^2}}Q(\omega) ; t} \notag \\
&= \frac{C_j}{m_j\omega_j^2}\Bigg\{ \prescript{W}{+}{\bm{D}_t^\alpha} Q(t) - \mathcal{F}^{-1}\squarebracket{(i\omega)^\alpha \frac{\omega^2}{\omega^2-\omega_j^2}Q(\omega) ; t}\Bigg\} \notag \\
&= \frac{C_j}{m_j\omega_j^2}\Bigg\{\prescript{W}{+}{\bm{D}_t^\alpha} Q(t) +\frac{d^2}{dt^2} \mathcal{F}^{-1}\squarebracket{(i\omega)^\alpha \frac{1}{\omega^2-\omega_j^2}Q(\omega) ; t}\Bigg\} \notag \\
&= \frac{C_j}{m_j\omega_j^2}\Bigg\{ \prescript{W}{+}{\bm{D}_t^\alpha} Q(t) +\frac{1}{2\pi}\frac{d^2}{dt^2}\mathcal{F}^{-1}\roundbracket{\frac{1}{\omega^2-\omega_j^2}; t}\ast \mathcal{F}^{-1}\squarebracket{(i\omega)^\alpha Q(\omega) ; t}\Bigg\}, 
\label{eq:middelstepeeee}
\end{align}
where in the second line we separated a term that will cancel the term coming from $\mathcal{L}_{CT}$ and in the last line we used the property of the inverse Fourier of a product in our conventions. We further simplify the above expression by making use of
\begin{align}
    \mathcal{F}^{-1}\roundbracket{\frac{1}{\omega^2-\omega_j^2}; t} &=-\frac{1}{2\omega_j}\Bigg[ \mathcal{F}^{-1}\roundbracket{\frac{1}{\omega + \omega_j}; t} -\mathcal{F}^{-1}\roundbracket{\frac{1}{\omega-\omega_j}; t}\Bigg] = -\frac{2\pi}{\omega_j} \sin(\omega_j t)\Theta(t),
    \label{eq:interm_appendix_}
\end{align}
where in the last line we interpreted the inverse Fourier integrals in the positive $i\epsilon$-interpretation. By the positive $i\epsilon$-interpretation, we mean that first we move the pole to the upper half of the complex plane, see Fig.~\ref{fig:contour_first}. After, we close the contour from above (below) for positive (negative) times. This allows one to use the residue theorem and Jordan's lemma to yield a finite value for these integrals. Finally, we put the added imaginary part to zero. This results in $\mathcal{F}^{-1}\squarebracket{1/(\omega - b); t} = \Theta(t)2\pi i e^{ibt}$, where $b$ is real. The positive $i\epsilon$-interpretation is justified, since in the limit of $\alpha \to 0$, we would like to recover the usual Caldeira-Leggett model. The physical justification is that the particle is only put in the bath after time $t=0$. Therefore, the interaction term should not have an impact before $t=0$. A Heaviside function, which automatically comes out of the positive $i\epsilon$-interpretation, ensures that, only after $t=0$, coupling between the bath and the particle occurs\footnote{The first term in Eq.~\eqref{eq:middelstepeeee} still shows a coupling between the bath and the particle, even for $t < 0$. However, this term will exactly cancel the term coming from $\mathcal{L}_{CT}$, such that it has no contribution in the equation of motion of the particle.}. Substituting Eq.~\eqref{eq:interm_appendix_} into Eq.~\eqref{eq:middelstepeeee} results in
\begin{align}
    q_j^p(t) &=\frac{C_j}{m_j\omega_j^2}\squarebracket{\prescript{W}{+}{\bm{D}_t^\alpha} Q(t) - \frac{1}{\omega_j}\frac{d^2}{dt^2} \sin (\omega_j t)\Theta(t)\ast \prescript{W}{+}{\bm{D}_t^\alpha} Q(t)}  \notag \\
    &= \frac{C_j}{m_j\omega_j^2}\Bigg\{ \prescript{W}{+}{\bm{D}_t^\alpha} Q(t) - \frac{d}{dt} \squarebracket{\cos (\omega_j t)\Theta(t) + \frac{1}{\omega_j}\sin(\omega_jt)\delta(t)} \ast \prescript{W}{+}{\bm{D}_t^\alpha} Q(t)\Bigg\}  \notag \\
    &= \frac{C_j}{m_j\omega_j^2}\squarebracket{\prescript{W}{+}{\bm{D}_t^\alpha} Q(t) - \frac{d}{dt} \cos (\omega_j t)\Theta(t)\ast \prescript{W}{+}{\bm{D}_t^\alpha} Q(t)} \notag \\
    &= \frac{C_j}{m_j\omega_j^2}\otherbracket{\prescript{W}{+}{\bm{D}_t^\alpha} Q(t) - \frac{d}{dt} \prescript{W}{+}{\bm{D}_t^\alpha} \squarebracket{\cos (\omega_j t)\Theta(t)\ast Q(t)}},
\label{eq:bathpreal_appendix}
\end{align}
\end{widetext}
where in the first line we acted with a derivative on $\sin(\omega_jt)\Theta(t)$, in the second line we used the fact that $\frac{d}{dt}\Theta(t) = \delta(t)$ which results in a term of the form $\delta(t)\sin(\omega_j t)\ast Q(t) \sim \sin(0) = 0$, and in the last line we applied Eq.~\ref{eq:property_weyl_convolution} to switch the order of the convolution and fractional derivative.


\section{The Mellin transform}
\label{app:mellin}

\begin{figure}[h]
\includegraphics[keepaspectratio=true,width=\columnwidth]{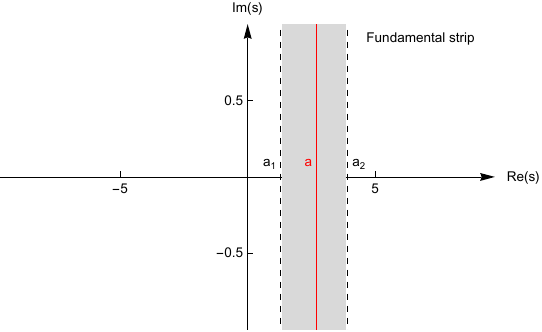}
    \caption{A depiction of the fundamental strip, needed for the inverse Mellin transform. Here, we chose $a_1 = 1.3$, $a = 2.7$ and $a_2 = 3.9$. This means that the fundamental strip in the figure is $S(1.3, 3.9)$.}
    \label{fig:inverse_mellin}
\end{figure}

The Mellin transform is defined by
\begin{align}
    \mathcal{M}\squarebracket{f(t);s} = \int_0^\infty dt f(t) t^{s-1}.
\end{align}
In general, the integral exists only for complex values of $s = a + ib$, such that $a_1 < a < a_2$, where $a_1$ and $a_2$ depend on the function $f(t)$ to transform. This introduces what is called the strip of definition of the Mellin transform, which will be denoted by $S(a_1, a_2)$. The Mellin transform of the exponential function is, e.g., given by
\begin{align}
    \mathcal{M}\squarebracket{e^{-at};s}
    &= a^{-s}\Gamma(s),
    \label{eq:mellin_exp}
\end{align}
which converges even for complex values of $a$ \cite{bateman} and Re$(s)>0$. Investigation of the Mellin transform of a more difficult function, namely $1/\roundbracket{1 + ct^a}$, is done by starting with the following equation
\begin{align}
    \int_0^\infty du \frac{u^{\frac{s}{a}-1}}{1+u} = \Gamma\roundbracket{\frac{s}{a}}\Gamma\roundbracket{1-\frac{s}{a}},
\end{align}
which can be derived from the beta function. Next, we apply the substitution $u = c t^a$ on the left-hand side, which results in\footnote{Note that the modulus is due to the reversal in the direction of integration for a negative power.} 
\begin{align}
    \int_0^\infty dt\frac{t^{s-1}}{1+ct^a} = \frac{1}{|a|}c^{-\frac{s}{a}}\Gamma\roundbracket{\frac{s}{a}}\Gamma\roundbracket{1-\frac{s}{a}}.
    \label{eq:mellin_examp}
\end{align}
Finally, we can identify the Mellin transform of the function $f(t)$ on the left-hand side of equation \eqref{eq:mellin_examp}, to conclude that
\begin{align}
\mathcal{M}\squarebracket{\frac{1}{1 + ct^a};s} =  \frac{1}{|a|}c^{-\frac{s}{a}}\Gamma\roundbracket{\frac{s}{a}}\Gamma\roundbracket{1-\frac{s}{a}},
\label{eq:mellin_exampleee}
\end{align}
where the region of convergence is given by $0 < \text{Re}(s)/a < 1$, see Ref.~\cite{bateman}. The inverse Mellin transform of a function, $F(s)$, defined on a strip $S(a_1, a_2)$ (see Fig.~\ref{fig:inverse_mellin}), is given by
\begin{align}
    f(t) = \mathcal{M}^{-1}\squarebracket{F(s); t} = \frac{1}{2\pi i} \int_{a-i\infty}^{a+i\infty} ds F(s)t^{-s},
    \label{eq:inverse_mellin}
\end{align}
where $a_1 < a < a_2$.

\section{The Fox H-function}
\label{app:foxh}

In this appendix, we briefly introduce the Fox H-function, see Refs. \cite{foxH, kilbas1999, foxhasympto}. The usefulness of the Fox H-function is that there are known asymptotic expansions. Therefore, when we derive an equality between Eq. \eqref{eq:fourier_int_def} and a Fox H-function, we also find the asymptotic expansions of the original integral.

Let $m, n, p, q \in \mathbb{N}$ with $0 \leq n \leq p$ and $1\leq m \leq q$, $A_i, B_j \in \mathbb{R}_{+}$, $a_i, b_j \in \mathbb{C}$ for $i = 1, ..., p$ and $j=1,..., q$. The Fox H-function with parameters $m, n, p, q,a_i, A_i, b_j ,B_j$ and argument $z\neq 0$ is defined by
    \begin{align}
        H^{m,n}_{p,q} \squarebracket{ z \middle| \begin{array}{ccc}
  \roundbracket{a_1,A_1},&...,&\roundbracket{a_p,A_p}\\
  \roundbracket{b_1,B_1},&...,&\roundbracket{b_q,B_q}
\end{array}} = \frac{1}{2\pi i} \int_L ds \Upsilon(s) z^{-s},
\label{eq:foxh_def}
\end{align}
where $z^{-s}$ is not necessarily the principal value complex power function and $\Upsilon(s)$ is given by
\begin{align}
    \Upsilon(s) = \frac{\prod_{i=1}^m\Gamma(b_i+sB_i)\prod_{l=1}^n\Gamma(1-a_l-sA_l)}{\prod_{i'=m+1}^q\Gamma(1-b_{i'}-sB_{i'})\prod_{l'=n+1}^p\Gamma(a_{l'}+sA_{l'})}.
    \label{eq:foxh_factor}
\end{align}
Here, an empty product is always interpreted as unity and $L$ is a suitable contour, which separates the poles of the two factors in the numerator of $\Upsilon(s)$. For more information about the contour, $L$, see Ref.~\cite{foxH}. 
We will also denote the Fox H-function as
\begin{align}
    H^{m,n}_{p,q}(z) \equiv H^{m,n}_{p,q} \squarebracket{ z \middle| \begin{array}{ccc}
  \roundbracket{a_1,A_1},&...,&\roundbracket{a_p,A_p}\\
  \roundbracket{b_1,B_1},&...,&\roundbracket{b_q,B_q}
\end{array}}
\end{align}
when the specification of the parameters, $a_l, A_l, b_i, B_i$, is not necessary. Let us now consider a simple example, namely, how the exponential function can be expressed in terms of the Fox H-function. We start with
\begin{align}
    H^{1,0}_{0,1} \squarebracket{ z \middle| \begin{array}{c}
  \\
  \roundbracket{0,1}
\end{array}} &= \frac{1}{2\pi i}\int_L ds \Gamma(s)z^{-s}.
\end{align}
Now, we choose the contour L to be the vertical line, $\text{\normalfont Re}(s) = \gamma$, $\gamma >0$, which lies to the right of the poles of $\Gamma(z)$. Thus, we have
\begin{align}
    H^{1,0}_{0,1} \squarebracket{ z \middle| \begin{array}{c}
  \\
  \roundbracket{0,1}
\end{array}} &=\frac{1}{2\pi i}\int_{\gamma -i\infty}^{\gamma + i\infty} ds \Gamma(s)z^{-s}.
\end{align}
Next, we assume that we can close the contour such that it picks up all the poles of the gamma function\footnote{A more formal treatment of this example can be found in Ref.~\cite{fox_h_examples}. In the same reference, more advanced examples are presented.}. Applying the residue theorem to the poles of the gamma function yields
\begin{align}
     H^{1,0}_{0,1} \squarebracket{ z \middle| \begin{array}{c}
  \\
  \roundbracket{0,1}
\end{array}} &= \sum_{\nu = 0}^\infty z^\nu\frac{(-1)^\nu}{\nu!}, \notag \\
&= e^{-z},
\end{align}
which shows that
\begin{align}
    e^{-z} = H^{1,0}_{0,1} \squarebracket{ z \middle| \begin{array}{c}
  \\
  \roundbracket{0,1}
\end{array}}.
\end{align}

Some quantities that will be important to determine the asymptotic behavior of Fox-H functions are defined as
\begin{align}
    &\mu_H \equiv \sum_{j=1}^qB_j-\sum_{j=1}^pA_j,\label{eq:foxh_mu}\\
    &\delta_H \equiv \sum_{j=1}^qb_j-\sum_{j=1}^pa_j + \frac{p-q}{2},\label{eq:foxh_delta}\\
    &\alpha_H \equiv \sum_{j=1}^nA_j - \sum_{j=n+1}^pA_j + \sum_{j=1}^mB_j - \sum_{j=m+1}^q B_j.
    \label{eq:foxh_alpha}
\end{align}
Now, we can use Section 1.7 from Ref.~\cite{foxH}, which shows the following properties.
\begin{enumerate}[label=(\roman*)]
\item If $\alpha_H > 0$ and $\left| \text{\normalfont arg}(z) \right|<\frac{1}{2}\pi \alpha_H$, then the Fox H-function has the asymptotic expansions at zero given by
\begin{align}
    H^{m,n}_{p,q}(z) &= \mathcal{O}(z^c), &|z| \to 0  \label{eq:short_time_foxh_theorem1}
\end{align}
for $\mu_H \geq 0$ and
\begin{align}
H^{m,n}_{p,q}(z) &= \mathcal{O}\roundbracket{z^c\left|\ln(z) \right|^{N-1}}, &|z|\to 0
    \label{eq:short_time_foxh_theorem2}
\end{align}
 for $\mu_H <0$, with $c = \min_{1\leq i \leq m}\squarebracket{\text{\normalfont Re}(b_i)/B_i}$ and $N$ is the order of the poles of $\Gamma(b_i+sB_i)$, to which some other poles of another gamma function $\Gamma(b_k+B_ks)$ with $1\leq i \leq m$ and $1\leq k \leq m$ could coincide.
\item If $\mu_H < 0$, $\alpha_H = 0$, then the asymptotic expansion near zero becomes
\begin{align}
    &H^{m,n}_{p,q}(z) = \mathcal{O}(z^\sigma), \qquad |z|\to 0,\: |\text{\normalfont arg}(z)|\leq \widetilde{\epsilon},
\end{align}
where $\widetilde{\epsilon}$ is a constant with\\ $0 < \widetilde{\epsilon} < (\pi/2)\min_{1\leq i \leq m; m+1\leq k \leq q}\roundbracket{A_i, B_k}$ and
$\sigma = \min_{1\leq i \leq m}\big\{\text{\normalfont Re}(b_i)/B_i, [\text{\normalfont Re}(\delta_H) + 1/2]/\mu_H\big\}$.
\item If $\alpha_H >0$, then the Fox H-function has the asymptotic expansion at infinity given by
\begin{align}
    H^{m,n}_{p,q}(z) &= \mathcal{O}(z^d), &|z|\to \infty \label{eq:long_time_foxh_theorem1}
\end{align}
for $\mu_H \leq 0$ and
\begin{align}
    H^{m,n}_{p,q}(z) &= \mathcal{O}(z^d\left| \ln (z) \right|^{M-1}), &|z|\to \infty
     \label{eq:long_time_foxh_theorem2}
\end{align}
for $\mu_H >0$, with $d = \min_{1\leq l \leq n}\squarebracket{(\text{\normalfont Re}(a_l)-1)/A_l}$ and $M$ is the order of the poles $\Gamma\roundbracket{1-a_l-sA_l}$ to which some of the poles of $\Gamma\roundbracket{1-a_k-sA_k}$ with $1\leq l \leq n$ and $1\leq k \leq n$ could coincide.
\item If $\mu_H >0$ and $\alpha_H =0$, then
\begin{align}
    &H^{m,n}_{p,q}(z) = \mathcal{O}(z^\rho), \qquad |z|\to \infty,\: \left|\text{\normalfont arg}(z) \right| \leq \epsilon,
\end{align}
where $\epsilon$ is a constant with \\
$0 < \epsilon < (\pi/2)\min_{n+1\leq j \leq p; 1\leq k \leq m}\roundbracket{A_j, B_k}$ and \\
$\rho = \max_{1\leq l \leq n}\big\{[\text{\normalfont Re}(a_l)-1]/A_l, [\text{\normalfont Re}(\delta_H)+1/2]/\mu_H\big\}$.
\end{enumerate}


\section{Asymptotic expansion of the MSD}
\label{app:asymptotic_msd}

Taking the Mellin transform of $I'(t)$, see Eq.~\eqref{eq:integeral_t}, we get
\begin{align}
    F(s) &\equiv \mathcal{M}\squarebracket{I'(t); s} \notag \\
    &= \int_0^\infty dt t^{s-1}\int_{-\infty}^\infty d\omega e^{i\omega t}\frac{(i\omega)^\alpha}{i\eta|\omega|^{2\alpha} - M\omega} \notag \\
    &= \frac{1}{M}\Gamma(s) \int_{-\infty}^\infty d\omega \frac{\roundbracket{i\omega}^\alpha \roundbracket{-i\omega}^{-s}}{i\frac{\eta}{M}|\omega|^{2\alpha} - \omega},
\end{align}
where in the last line we used Eq.~\ref{eq:mellin_exp}. This already gives a restriction on $s$, namely $\text{Re}(s) > 0$. Next, we apply properties of the complex power function, together with the fact that $\omega \in \mathbb{R}$, which results in
\begin{align}
    F(s) &= \frac{1}{M}\Gamma(s) \int_{-\infty}^\infty d\omega \frac{|\omega|^{\alpha-s}e^{\text{arg}(i\omega)i\alpha - \text{arg}(-i\omega)is}}{i\frac{\eta}{M}|\omega|^{2\alpha} - \omega} \notag \\
    &= \frac{1}{M}\Gamma(s) \int_{-\infty}^\infty d\omega \frac{|\omega|^{\alpha-s}e^{i\text{arg}(i\omega)\roundbracket{\alpha + s}}}{i\frac{\eta}{M}|\omega|^{2\alpha} - \omega}.
\end{align}
We can further simplify the above equation by splitting the integral in its positive and negative parts. Doing the substitution $\omega \to -\omega$ in the latter yields
\begin{widetext}
    \begin{align}
    F(s) &=  \frac{1}{M}\Gamma(s)\squarebracket{e^{i\frac{\pi}{2}\roundbracket{\alpha + s}}\int_0^\infty d\omega \frac{\omega^{\alpha-s}}{i\frac{\eta}{M}\omega^{2\alpha} - \omega} + e^{-i\frac{\pi}{2}\roundbracket{\alpha + s}}\int_{0}^\infty d\omega \frac{\omega^{\alpha-s}}{i\frac{\eta}{M}\omega^{2\alpha} + \omega}} \notag \\
    &= \frac{1}{M}\Gamma(s)\squarebracket{e^{-i\frac{\pi}{2}\roundbracket{\alpha + s}}\int_{0}^\infty d\omega \frac{\omega^{\alpha-s-1}}{1+i\frac{\eta}{M}\omega^{2\alpha-1}} - e^{i\frac{\pi}{2}\roundbracket{\alpha + s}}\int_{0}^\infty d\omega \frac{\omega^{\alpha-s-1}}{1-i\frac{\eta}{M}\omega^{2\alpha - 1}}} \notag \\
    &= \frac{1}{M|2\alpha-1|}\squarebracket{e^{-i\frac{\pi}{2}(\alpha + s)}\left(i\frac{\eta}{M}\right)^{\frac{s-\alpha}{2\alpha-1}}- e^{i\frac{\pi}{2}(\alpha +s)}\left(-i\frac{\eta}{M}\right)^{\frac{s-\alpha}{2\alpha-1}}} \Gamma(s)\Gamma\roundbracket{\frac{\alpha-s}{2\alpha-1}}\Gamma\roundbracket{1-\frac{\alpha-s}{2\alpha-1}} \notag \\
    &= \frac{1}{M|2\alpha-1|}\roundbracket{\frac{\eta}{M}}^{\frac{s-\alpha}{2\alpha-1}}\squarebracket{e^{-i\frac{\pi}{2}\left(s + \frac{2\alpha^2}{2\alpha - 1}\right)}- e^{i\frac{\pi}{2}\left(s + \frac{2\alpha^2}{2\alpha - 1}\right)}} \Gamma(s)\Gamma\roundbracket{\frac{\alpha-s}{2\alpha-1}}\Gamma\roundbracket{1-\frac{\alpha-s}{2\alpha-1}} \notag \\
    &= -\frac{2i}{M|2\alpha-1|}t_s^{s-\alpha}\sin\squarebracket{\frac{\pi}{2}\left(s + \frac{2\alpha^2}{2\alpha - 1}\right)}\Gamma(s)\Gamma\roundbracket{\frac{\alpha-s}{2\alpha-1}}\Gamma\roundbracket{1-\frac{\alpha-s}{2\alpha-1}},
\end{align}
where in the third line we applied Eq.~\eqref{eq:mellin_exampleee} to each term separately and in the last line we identified the characteristic time-scale of the system given by $t_s = (M/\eta)^{1/(1-2\alpha)}$. However, using this example restricts the values of $s$ even further. The additional constraint reads $0 < \text{Re}\roundbracket{\frac{\alpha - s}{2\alpha - 1}} < 1$. Together with the previous constraint that we already mentioned, $\text{Re}(s) > 0$, we obtain three different regions of convergence that are important to define the inverse Mellin transform, see Eq.~\eqref{eq:inverse_mellin}. Just like in the definition, we call them fundamental strips (FS) and they are given by
\begin{equation}
\text{FS} = 
\begin{cases}
    &S(0, \alpha)\qquad \ \ \ \ \ \ \text{ if } \alpha > 1,  \\
    &S(1-\alpha, \alpha)\qquad \: \text{ if } \frac{1}{2} < \alpha < 1, \\
    &S(0, 1-\alpha)\qquad \ \text{ if } \alpha < \frac{1}{2}.
\end{cases}
\label{eq:funda_strip}
\end{equation}
Finally, we use Euler's reflection formula to get 
\begin{align}
    F(s) = -\frac{2\pi i}{M|2\alpha-1|}t_s^{s-\alpha}\frac{\Gamma(s)\Gamma\roundbracket{\frac{\alpha-s}{2\alpha-1}}\Gamma\roundbracket{1-\frac{\alpha-s}{2\alpha-1}}}{\Gamma\roundbracket{\frac{\alpha^2}{2\alpha - 1} + \frac{s}{2}}\Gamma\roundbracket{1-\frac{\alpha^2}{2\alpha - 1} - \frac{s}{2}}},
    \label{eq:before_split_mellin_app}
\end{align}
which completes the derivation of Eq.~\eqref{eq:before_split_mellin}.
In the case of the first and second fundamental strip, where $2\alpha -1 > 0$, we rewrite Eq.~\eqref{eq:before_split_mellin_app} as
\begin{align}
    F(s) = -\frac{2\pi i}{M|2\alpha-1|}t_s^{s-\alpha}\frac{\Gamma(s)\Gamma\roundbracket{1- \frac{\alpha-1}{2\alpha -1} - \frac{s}{2\alpha - 1}}\Gamma\roundbracket{\frac{\alpha-1}{2\alpha -1} + \frac{s}{2\alpha -1}}
    }{\Gamma\roundbracket{\frac{\alpha^2}{2\alpha - 1} + \frac{s}{2}}\Gamma\roundbracket{1-\frac{\alpha^2}{2\alpha - 1} - \frac{s}{2}}}.
    \label{eq:intermediateddfe}
\end{align}
Finally, we take the inverse Mellin transform, defined by Eq.~\eqref{eq:inverse_mellin}, which results in
\begin{align}
    I'(t) &= \mathcal{M}^{-1}\squarebracket{F(s);t} = \frac{1}{2\pi i} \int_{a-i \infty}^{a + i \infty} ds F(s)t^{-s},
\end{align}
where $F(s)$ is given by Eq. \eqref{eq:intermediateddfe}. Comparing the above equation with Eqs.~\eqref{eq:foxh_def}~and~\eqref{eq:foxh_factor}, we get 
\begin{align}
    I'(t) = -\frac{2\pi i}{M|2\alpha-1|}t_s^{-\alpha}
 H^{21}_{23} \left[ \frac{t}{t_s}\ \middle| \begin{array}{ccc}
  & \left( \frac{\alpha-1}{2\alpha -1}, \frac{1}{2\alpha - 1}\right), & \left( \frac{\alpha^2}{2\alpha - 1}, \frac{1}{2}\right)\\
  (0, 1) & \left( \frac{\alpha - 1}{2\alpha - 1}, \frac{1}{2\alpha - 1} \right), & \left( \frac{\alpha^2}{2\alpha -1}, \frac{1}{2}\right) 
\end{array} \right],
\label{eq:fox_firstfunda}
\end{align}
which is exactly Eq.~\eqref{eq:foxH_final}. Note that Eq.~\eqref{eq:fox_firstfunda} holds for both the first and second fundamental strips. Next, we focus on the third fundamental strip, and rewrite Eq.~\eqref{eq:before_split_mellin_app} as
\begin{align}
   F(s) = -\frac{2\pi i}{M|2\alpha-1|}t_s^{s-\alpha} \frac{\Gamma(s)\Gamma\roundbracket{\frac{\alpha}{2\alpha-1}+\frac{s}{1-2\alpha}}\Gamma\roundbracket{1-\frac{\alpha}{2\alpha-1} - \frac{s}{1-2\alpha}}}{\Gamma\roundbracket{\frac{\alpha^2}{2\alpha - 1} + \frac{s}{2}}\Gamma\roundbracket{1-\frac{\alpha^2}{2\alpha - 1} - \frac{s}{2}}},
\end{align}
which, when taking the inverse Mellin transform and comparing with Eqs.~\eqref{eq:foxh_def}~and~\eqref{eq:foxh_factor} results in 
\begin{align}
    I'(t) &= -\frac{2\pi i}{M|2\alpha-1|}t_s^{-\alpha} H^{21}_{23} \left[\frac{t}{t_s}\ \middle| \begin{array}{ccc}
  & \left( \frac{\alpha}{2\alpha-1}, \frac{1}{1-2\alpha}\right), & \left( \frac{\alpha^2}{2\alpha-1}, \frac{1}{2}\right)\\
  (0, 1) & \left(\frac{\alpha}{2\alpha-1}, \frac{1}{1-2\alpha} \right), & \left( \frac{\alpha^2}{2\alpha-1}, \frac{1}{2}\right) 
\end{array} \right].
\label{eq:foxH_final2}
\end{align}
\end{widetext}
 What remains is to apply the known asymptotic expansions for the Fox H-function to Eqs.~\eqref{eq:fox_firstfunda}~and~\eqref{eq:foxH_final2}. First, we note that in both cases we have $\mu_H, \alpha_H > 0$. This means that the short and long time expansion are given by Eqs.~\eqref{eq:short_time_foxh_theorem1} and~\eqref{eq:long_time_foxh_theorem2}, respectively. The short-time behavior is given by
\begin{align}
    I'(t) \stackrel{t \to 0}{\sim} \begin{cases}
&1 \qquad \quad \ \ \ \text{ for } \alpha>1, \\
&t^{\alpha -1} \qquad \ \: \text{ for } \frac{1}{2}<\alpha < 1,\\
&t^{\frac{2\alpha^2}{2\alpha - 1}} \qquad \: \text{ for }\frac{1}{4} < \alpha < \frac{1}{2},\\
&t^{-\alpha} \qquad \ \ \ \text{ for }0 < \alpha \leq \frac{1}{4},\\
&t^{\frac{2\alpha^2}{2\alpha - 1}} \qquad \ \text{ for }\alpha < 0,\\
\end{cases}
\label{eq:firstexpans}
\end{align}
and the long-time behavior is
\begin{align}
    I'(t) \stackrel{t \to \infty}{\sim} \begin{cases}
&t^{-\alpha} \quad \quad \ \: \text{ for } \alpha > 1, \\
&t^{\alpha-1} \quad \: \quad \text{ for } \alpha < 1, \alpha \neq 0.
\end{cases}
\label{eq:secondexpans}
\end{align}
From now on we will discard the case where $\alpha = -1/2, (-1\pm \sqrt{3})/2, (-3 \pm \sqrt{21})/4$. This is because we want to keep the calculation as simple as possible and therefore do not want any logarithms while integrating the expansions. We thus have that $\alpha \in \mathbb{R}\setminus \mathcal{P}$ with $\mathcal{P} = \big\{(-3 - \sqrt{21})/4, (-1 - \sqrt{3})/2, -1/2, 0,(1 + \sqrt{3})/2,(-3 + \sqrt{21})/4, 1/2, 1 \big\}$. Substituting the expansions given by Eqs.~\eqref{eq:firstexpans}~and~\eqref{eq:secondexpans} into Eq.~\eqref{eq:integeral_t} adds a one to the exponents of the asymptotic expansions due to the integration\footnote{To calculate the integral over the small time asymptotic expansion of $I'(t)$ which results in the asymptotic expansions of $I(t)$, one has to technically introduce a small time cut-off since otherwise the evaluation at $t = 0$ would diverge.}. Thus, we obtain 
\begin{align}
    I(t) \stackrel{t \to 0}{\sim} \begin{cases}
&t \qquad \quad \ \ \: \: \quad \text{ for } \alpha>1, \\
&t^{\alpha} \qquad \ \: \qquad \text{ for } \frac{1}{2}<\alpha < 1,\\
&t^{\frac{2\alpha^2}{2\alpha - 1} + 1} \qquad \: \text{ for }\frac{1}{4} < \alpha < \frac{1}{2},\\
&t^{-\alpha + 1} \qquad \ \ \ \text{ for }0 < \alpha \leq \frac{1}{4},\\
&t^{\frac{2\alpha^2}{2\alpha - 1} + 1} \qquad \ \text{ for }\alpha < 0,\\
\end{cases}
\end{align}
where $\alpha \notin \mathcal{P}$. Finally, we can substitute these asymptotic expansions into Eq.~\eqref{eq:solution_frac_lang} and Taylor expand the exponential factor up to first order, to get the small time behavior of the MSD, which is given by
\begin{align}
    \left\langle Q(t)^2 \right\rangle \stackrel{t \to 0}{\sim} \begin{cases}
        &t^2 \qquad \text{ for } \alpha > \frac{-1 - \sqrt{5}}{4}, \alpha \notin \mathcal{P},\\
        &t^{2\alpha + 3} \quad \text{for } \alpha < \frac{-1-\sqrt{5}}{4}, \alpha \notin \mathcal{P}.
        \end{cases}
\label{eq:shorttime_msd}
\end{align}
We can do exactly the same for the long-time behavior, except that now we discard the exponential in Eq.~\eqref{eq:solution_frac_lang} because it vanishes as $t \to \infty$. The long-time behavior of the MSD is
\begin{align}
    \left\langle Q(t)^2 \right\rangle \stackrel{t \to \infty}{\sim} \begin{cases}
&t^{-2\alpha + 3} \quad \ \text{ for } \alpha > 1, \alpha \notin \mathcal{P},\\
&t^{2\alpha + 1}\quad  \ \ \ \text{ for } \alpha < 1, \alpha \notin \mathcal{P}.
\end{cases}
\label{eq:long_time_msd_final}
\end{align}

\bibliography{ref}

\end{document}